\documentclass[lettersize,journal]{IEEEtran}
\usepackage{times}
\usepackage{amsmath,amsfonts}
\usepackage{algorithmic}
\usepackage{algorithm}
\usepackage{array}
\usepackage[caption=false,font=normalsize,labelfont=sf,textfont=sf]{subfig}
\usepackage{textcomp}
\usepackage{stfloats}
\usepackage{url}
\usepackage{verbatim}

\usepackage{cite}
\usepackage{graphicx}
\usepackage[hidelinks]{hyperref}
\graphicspath{{figs/}}
\usepackage{xcolor}

\usepackage{makecell}

\usepackage{enumitem}
\usepackage{multirow}
\usepackage{amssymb}
\usepackage{newtxtext, newtxmath}
\usepackage{amsmath}
\usepackage{diagbox}

\begin{document}

\title{CWRNN-INVR: A Coupled WarpRNN based Implicit Neural Video Representation}

\author{Yiyang Li\textsuperscript{*}, Yanbo Gao\textsuperscript{*}, Shuai Li\textsuperscript{$\dagger$},~\IEEEmembership{Senior Member,~IEEE}, Zhenyu Du, Jinglin Zhang, \\Hui Yuan,~\IEEEmembership{Senior Member,~IEEE}, Mao Ye,~\IEEEmembership{Senior Member,~IEEE}, Xingyu Gao

	\thanks{Y. Li, S. Li, Z. Du, J. Zhang and H. Yuan are with School of Control Science and Engineering, Shandong University, and Key Laboratory of Machine Intelligence and System Control, Ministry of Education, Jinan 250100, China. E-mail:  shuaili@sdu.edu.cn. \par Y. Gao is with School of Software, Shandong University, Jinan 250100, China, and also with Shandong University-WeiHai Research Institute of Industrial Technology, Weihai 264209, China. \par M. Ye is with University of Electronic Science and Technology of China, Sichuan, China. E-mail: cvlab.uestc@gmail.com  \par X. Gao is with the Institute of Microelectronics, Chinese Academy of Sciences, Beijing 100029, China 
}}



\maketitle

\begin{abstract}
Implicit Neural Video Representation (INVR) has emerged as a novel approach for video representation and compression, using learnable grids and neural networks. Existing methods focus on developing new grid structures efficient for latent representation and neural network architectures with large representation capability, lacking the study on their roles in video representation. In this paper, the difference between INVR based on neural network and INVR based on grid is first investigated from the perspective of video information composition to specify their own advantages, i.e., neural network for general structure while grid for specific detail. Accordingly, an INVR based on mixed neural network and residual grid framework is proposed, where the neural network is used to represent the regular and structured information and the residual grid is used to represent the remaining irregular information in a video.  A Coupled WarpRNN-based multi-scale motion representation and compensation module is specifically designed to explicitly represent the regular and structured information, thus terming our method as CWRNN-INVR. For the irregular information, a mixed residual grid is learned where the irregular appearance and motion information are represented together. The mixed residual grid can be combined with the coupled WarpRNN in a way that allows for network reuse. Experiments show that our method achieves the best reconstruction results compared with the existing methods, with an average PSNR of 33.73 dB on the UVG dataset under the 3M model and outperforms existing INVR methods in other downstream tasks. The code can be found at \href{https://github.com/yiyang-sdu/CWRNN-INVR.git}{https://github.com/yiyang-sdu/CWRNN-INVR.git.}
\end{abstract}


\begin{figure}[t]
    \centering
    \includegraphics[width=0.35\textwidth]{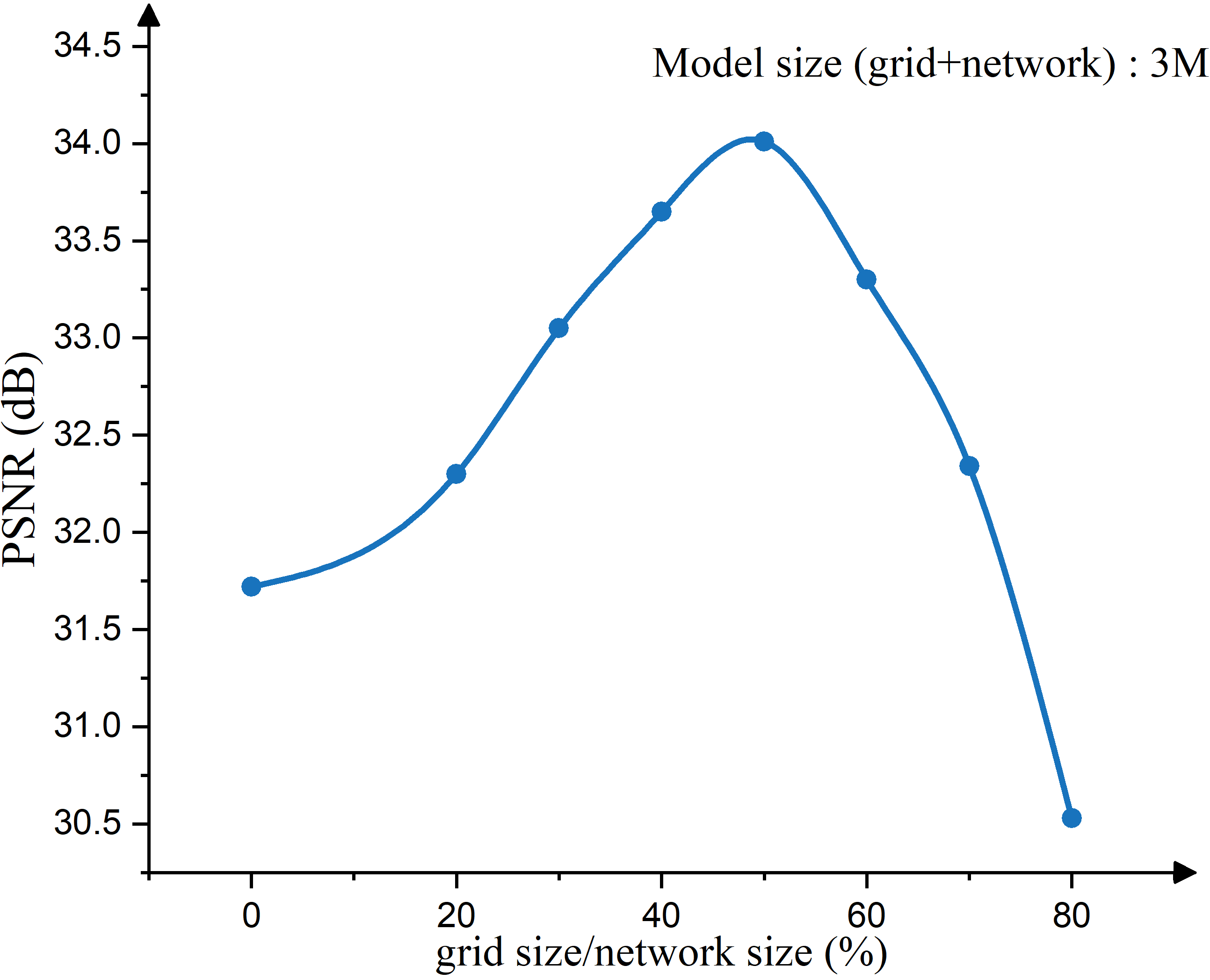}
    \caption{Illustration of using different combinations of grids and neural networks (only NeRVBlocks) while keeping the size of the total model constant, for video reconstruction on the Big Buck Bunny dataset. Performance varies significantly with different combinations, indicating that grids and networks each have their own advantages in terms of feature representation.}
    \label{Fig. 1}
\end{figure}

\section{Introduction}
\label{intro}
\IEEEPARstart{I}{mplicit}  Neural Representations (INRs) \cite{mildenhall2021nerf,chen2021nerv,sitzmann2020implicit,chen2022tensorf,muller2022instant,barron2021mip} have gained increasing interests recently due to its strong ability in learning feature representations. Among the various signal types represented by INR, Implicit Neural Video Representation (INVR) \cite{chen2021nerv,chen2023hnerv,he2023towards,li2022nerv,maiya2023nirvana,lee2023ffnerv}  to fit a video has been widely studied. Existing INVR methods \cite{bai2023ps,gupta2024pnerv,yan2024ds,zhao2024pnerv,kwan2024hinerv} can be broadly categorized into two approaches considering their 
representation formats. The first approach takes the frame index as input and uses MLP or other neural networks to reconstruct the video frame at each frame index. This approach focuses on developing efficient neural networks to represent videos, and can be termed as Implicit Neural Video Representations based on Neural Network (INVR-N), as in \cite{chen2021nerv,li2022nerv}. The other approach uses a set or multiple sets of {learnable variables/embeddings, such as grids or codes}, as the 2D video frame features, which are then processed by NeRVBlocks to reconstruct frames \cite{chen2023hnerv,lee2023ffnerv,yan2024ds}. Taking HNeRV as an example, it uses an encoder to explicitly produce 
{a latent embedding} as input for the representation. This approach focuses on constructing efficient grid representations while only using the neural network as a decoder, and can be referred to as Implicit Neural Video Representation based on Grid (INVR-G)\cite{gop}. However, the differences between using neural networks and grids have not been thoroughly investigated, and the benefits of both representations are not fully explored yet. 

From the perspective of representation, a video sequence actually contains two types of information. One is the structured information with regular patterns that can be easily represented by a neural network in the spatiotemporal domain, including structured appearance and structured motion that regularly exist across frames \cite{9899414,10549826,10891391,9328173}. This kind of information can be well learned by neural networks. The other is irregular information with patterns hard to be represented by a neural network, including unique local appearance and varying motion that only irregularly exist in one or serval frames. Such information cannot be efficiently represented by a neural network since they only occur occasionally in a video. In such a case, a specifically learned grid can be used to represent the irregular information with a small representation cost.

\begin{figure*}[t]
  \centering
    \includegraphics[width=0.95\textwidth]{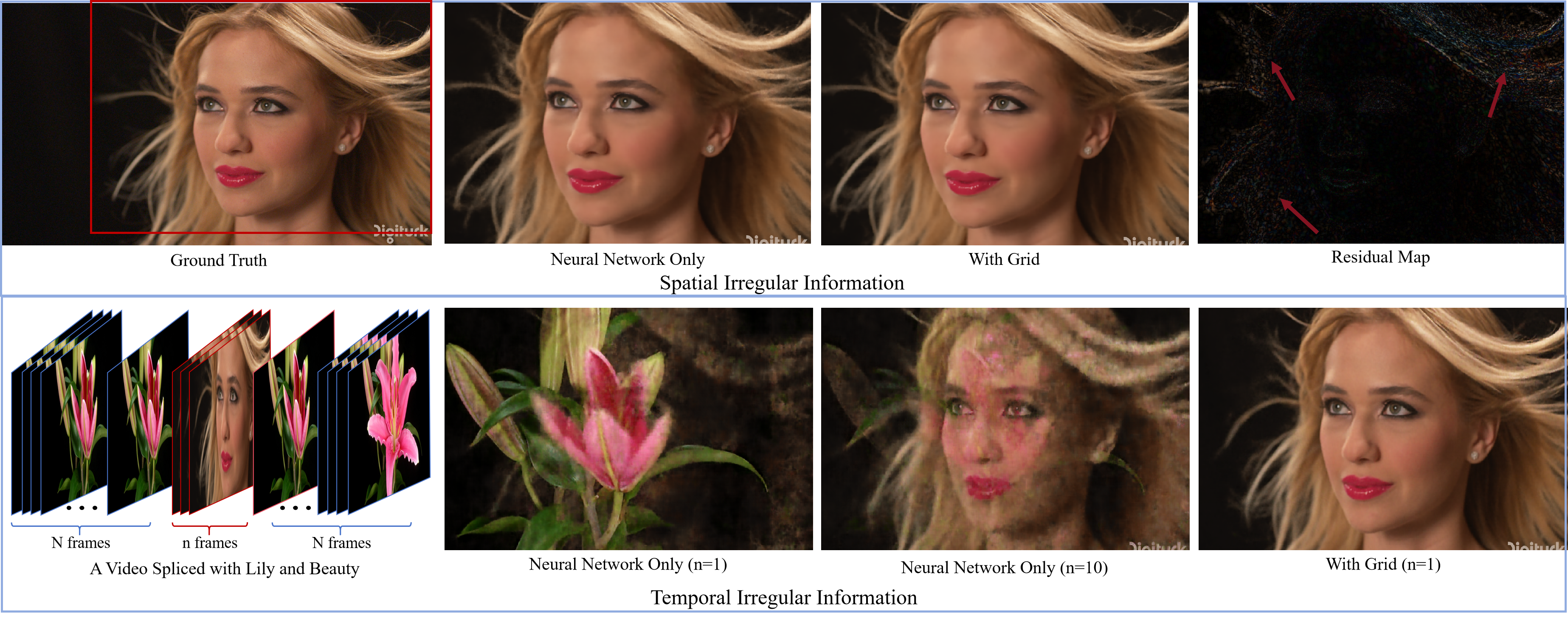}
    \caption{Illustration of the spatial and temporal regular and irregular information in a video. The top shows the comparison of the reconstruction results using neural network only (without grid) and with grid (under the same number of parameters in total) on the UVG Beauty video. The residual map is obtained by subtracting the two results. The bottom is the reconstruction results on a spliced video that n frames of the Beauty are inserted between the 2N frames of a blooming lily video. The residual map in the top illustrates the spatial irregular information, mainly on the hair which is hard to be represented and not structured among video frames. The bottom results using the spliced video clearly show the temporal irregular information, which is difficult to be represented by a neural network but easily represented by a grid.  }
    \label{motivation}
\end{figure*}

Fig. \ref{Fig. 1} illustrates using different combinations of grids and neural networks for video reconstruction on the Big Buck Bunny dataset \cite{roosendaal2008big} where the combination is measured by the percentage of grid parameters and network parameters (grid/network ratios). A smaller network leads to a large ratio and vice versa. It can be seen that performance varies significantly with different combinations, indicating that grids and networks each have their own advantages in terms of feature representation. By comparing a pair of large and small grid/network ratios, it can be seen that the performance dropped significantly, i.e., using a smaller neural network is much worse than using a smaller grid at a similar ratio, 32.02 dB versus 30.53 dB under ratios of 20\% and 80\%, respectively. This illustrates that neural network is more efficient in representing regular and structured information than grid since the overall quality is largely controlled by the regular information repeatedly occurred in a video. 

To further demonstrate the structured information with regular patterns  and irregular information, Fig. \ref{motivation} illustrates the reconstruction results using neural network only (without grid) and with grid (under the same number of parameters in total). By comparing the face (easy to represent and regularly occurred) and hair (difficult to represent and changed over frames) in the Beauty video, it can be seen that the model with Grid can better capture more spatial irregular information details than using neural network only. The residual map is further demonstrated in the right to show the differences. To better visualize the temporal irregular information, a video created by inserting n frames of the Beauty video between 2N frames of a blooming lily video is used. The results using neural network only (without grid) and with Grid are illustrated. It can be clearly seen that when only one frame is inserted ($n=1$), the frame is constructed mostly with the temporal regular information, which is the Lily flower in the video. Even when $n=10$, the temporal regular information is still visible in the reconstruction. On the other hand, when grid is used, the temporally irregular information can be corrected reconstructed even when $n=1$. This verifies the necessity using the neural network and grid to represent the regular and irregular information. The detailed experimental settings and  quantitative results are explained in the later experiment section.

On the other hand, the existing INVR methods focus more on using neural network to represent spatial appearance of each frame while lacking investigation on using neural network to represent temporal motion. With  global motion such as camera movement and local motion such as object change, it is difficult to use a single network for high-quality motion representation, especially under the constraint of network size.

Motivated by the above observations, an Implicit Neural Video Representation based on Mixed neural network and grid (INVR-M) is developed, where the neural network is used to hard-code the regular and structured information and the grid is used to represent the remaining irregular information. A WarpRNN architecture is first developed based on recurrent neural networks (RNN), where the hidden state is enhanced with a motion based warping to perform temporal alignment between frames. It leverages the capabilities of RNN in temporal modeling and enhances it with motion compensation. To better represent the complex motion in a neural network, a multi-scale motion compensation module is proposed based on a coupled WarpRNN, where the global and local motion are separately formulated and progressively compensated. The remaining irregular motion and appearance information are represented together with a mixed residual grid, considering they are interleaved with each other from the perspective of rate distortion optimization. The appearance and motion are then generated with different projections of the mixed residual. In summary, our contributions are as follows:

\begin{itemize}
\item[  1)]A new Implicit Neural Video Representation framework based on Mixed neural network and residual grid, termed INVR-M, is proposed. It decomposes a video into regular and structured information, which can be easily represented by a neural network, and remaining irregular information represented by a grid.  

\item[  2)]A coupled WarpRNN based multi-scale motion representation and compensation module is developed as the temporal representation network, based on global and local motion based decomposition.

\item[  3)]A mixed residual grid is developed to represent the irregular appearance and motion information, in order to fully explore the correlation between appearance and motion.

\item[  4)]Extensive experiments are conducted on video reconstruction and various downstream tasks, and our method shows the best performance, validating its effectiveness.
\end{itemize}

\section{Related Work}
\noindent \textbf{Implicit Neural Representation.} In recent years, implicit neural representations (INR) have emerged as a new approach for parameterizing 3D scenes \cite{song2023nerfplayer,fridovich2023k,pumarola2021d,wang2023neural,wang2024videorf} and 2D video signals \cite{chen2021nerv,chen2023hnerv,sitzmann2020implicit}. Neural Radiance Fields (NeRF) \cite{mildenhall2021nerf} introduced the concept of fitting a 3D scene into a MLP network, using coordinate to map to network values and reconstructing the 3D scene through volumetric rendering. Based on such implicit neural representation, Implicit Neural Video Representations (INVR) has also been studied. In \cite{sitzmann2020implicit}, a pixel-wise video reconstruction method was proposed to fit the frame index  \(t\) and coordinate values into a MLP network and decode the video by querying all pixel coordinates. The decoding speed of this method is relatively slow due to the pixel-wise generation. Then NeRV \cite{chen2021nerv} proposed a frame-wise video reconstruction method with a frame-by-frame query approach, significantly improving the decoding efficiency. ENeRV \cite{li2022nerv,xf1} further incorporates the spatial coordinates into enhance the frame index based generation by decomposing the spatial and temporal information. In these methods, all the video information is represented in the neural network while the frame number and spatial coordinates are only used as index to extract the corresponding information.

On the other hand, there have also been INVR methods using learnable grids as input. Such methods reconstruct a video using the combination of embeddings (grids or codes) and neural network. For example, HNeRV \cite{chen2023hnerv} proposed to use a encoder to generate embeddings for frames, and then decode the embeddings with a neural network to reconstruct the video. The neural network works as a decoder to decode the information contained in the embeddings to frames. Instead of using a embedding for each frame, DNeRV \cite{zhao2023dnerv} further proposed to use a content embedding and difference flows to encode the frames based on a main frame and the difference between frames. In the FFNeRV \cite{lee2023ffnerv}, multi-resolution temporal grid was proposed to represent information at different temporal frequencies. Optical flow is also extracted at the end to enhance the consistency between frames. DSNeRV \cite{yan2024ds} proposed to use two grids containing the static codes and dynamic codes to representing the static or slowly moving area and the rich dynamic area, respectively. These methods focus on designing efficient grid structures to represent the video while the network is mostly used for decoding the grid features.  {LatentINR \cite{maiya2024laten} also employs the learnable latent features as input. The learnable latent features are used to capture semantic information of video frames, and then aligned with features from video LLMs for downstream tasks.} However, the differences between using neural networks and grids have not been fully studied. Moreover, there has been no research on exploring the neural network architecture for explicit temporal modelling in a video yet. This paper investigates the different roles of grid and neural network and proposed a mixed representation based on the RNN architecture for explicit temporal modeling. 
\par\noindent\textbf{Video Compression. }In the field of video compression, traditional methods typically employ a hybrid video  coding architectures with intra/inter prediction to reduce the spatial and temporal redundancy in a video such as the High Efficiency Video Coding (HEVC) \cite{sullivan2012overview}, Versatile Video Coding (VVC) \cite{xf6,bross2021overview} and other methods \cite{chen2018overview,legall1993video,xf4}. Recently learned video compression methods \cite{li2021deep,agustsson2020scale,yang2020learning,xf2,xf3} based on deep learning are also rapidly developing \cite{hu2021fvc,chen2017deepcoder,rippel2021elf}. These methods \cite{chen2019learning,lin2020m,lu2019dvc} usually learn a general large neural network and compress videos with a similar workflow as the traditional methods but with learnable components \cite{10129217,10485621,10697103}. With the emergence of implicit neural representation methods, INVR has also been used for video compression, where the learned grids (if any) and neural network are encoded. In this way, the video compression problem is transformed into a model compression problem \cite{dupont2021coin,lee2023entropy,zhang2024efficient,yang2023tinc}. INVR based video compression achieves performance on par with the existing learned video compression methods, with much less decoding complexity \cite{10254316,10855451,9941493,xf5}. This paper also investigates using the proposed CWRNN- INVR for video compression and achieves state-of-the-art compression performance.

\begin{figure*}[t]
  \centering
    \includegraphics[width=0.85\textwidth]{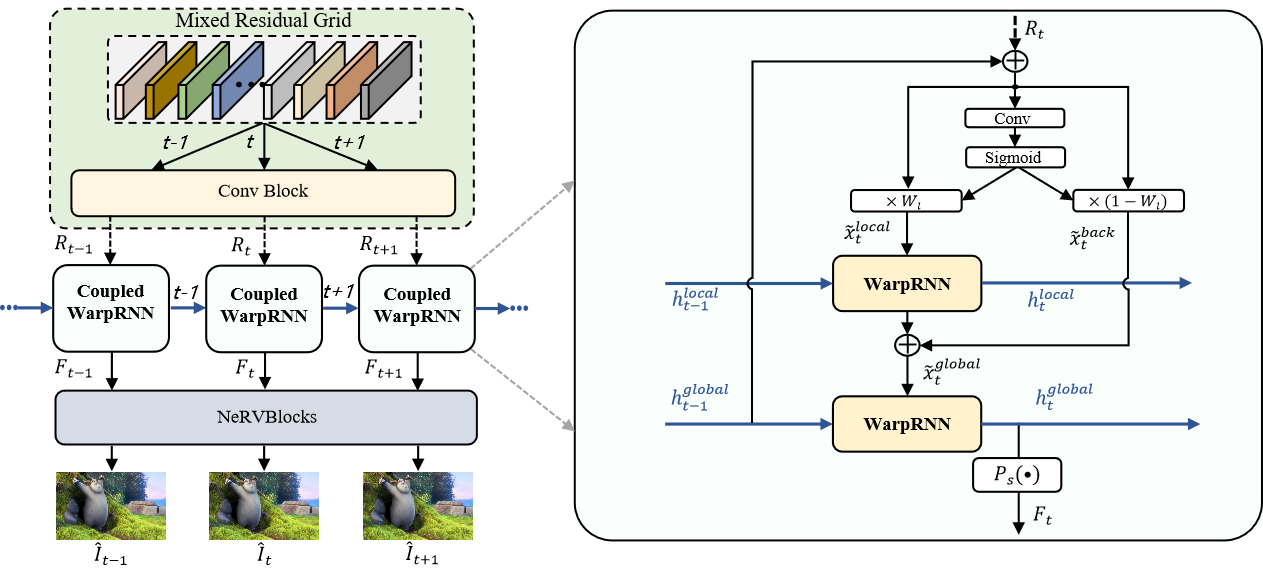}
    \caption{Left: Framework Overview. The proposed  Coupled WarpRNN based multi-scale motion representation and compensation module is used to learn the regular and structured temporal information and the following NeRVBlocks are used to spatially reconstruct the frame. It can work independently without grid or any input based on its recurrent architecture. The Mixed Residual Grid is used to learn the remaining irregular information and added to the coupled WarpRNN for network reuse.  Right: Structure of the proposed Coupled WarpRNN based multi-scale motion representation and compensation module.}
    \label{Fig. 2}
\end{figure*}

\section{Proposed Method}
\subsection{Formulation}
\label{sec3.1}
\noindent Implicit Neural Video Representation (INVR) uses learnable structures to represent a video, including neural networks and embeddings such as grid. However, the difference behind network and grid based representations are not fully investigated and their working mechanisms are unclear. In this paper, we propose an INVR-M framework based on mixed neural network and grid, which explicitly formulates the regular and structured information with a neural network and the irregular information with a grid. Given a video sequence with T frames, INVR-M formulates each frame by
\begin{equation}
  \hat{I}_t = F \oplus m = f_n(m),
  \label{Eq.(1)}
\end{equation}
where \( F \) denotes the neural network to represent the regular and structured spatial appearance and temporal motion information, and \( m \) denotes the mixed residual grid to represent the remaining irregular information. \( \oplus \) represents the combination of the network and mixed residua grid. To explore the network reuse, the mixed residual grid is also processed by the network. Thus it can also be represented as \( f_n(m) \) as shown in the above equation, where \( f_n(\cdot ) \) represents the processing of the network  \( F \). { The network as a function $f_n$ and the mixed residual grid $m$ together represent a video, and in the category of INVR based video compression, both the network and grid need be coded and transmitted to the user side.  

Regarding the network,} considering that a video not only contains spatial appearance information, but also temporal motion information, a spatial-temporal processing network is used in this paper. Moreover, in view that the complex motion contained in a video, a multi-scale motion prediction and compensation module is developed to formulate the local motion and global motion separately. Specifically, a coupled WarpRNN based architecture is developed to progressively perform the local and global motion prediction and compensation. For the spatial appearance processing, the NeRVBlocks are used for simplicity. This process can be represented as
\begin{equation}
  f_n = f_t \oplus f_s,
  \label{Eq.(2)}
\end{equation}
where \( f_t \) represents the coupled WarpRNN based multi-scale motion prediction and compensation module (as shown in Fig. \ref{Fig. 2}). \( f_s \) represents the NeRVBlocks to reconstruct the frame \(\hat{I}_t \) from the temporally compensated frame features.

On the other hand, the mixed residual grid represents the remaining irregular motion and appearance information that cannot be efficiently represented by the above spatial-temporal processing  network\cite{10736428,10814696}. Since the motion and appearance are interleaved with each other in the representation, a single mixed residual grid is used. The residual motion and residual appearance can be obtained with a temporal projection and a spatial projection, respectively, from the mixed grid. This process can be represented by
\begin{equation}
  [m_t, m_s] = [P_t, P_s](m),
  \label{Eq.(3)}
\end{equation}
where \( m_t\) and \( m_s\) represent the irregular motion and appearance obtained via the temporal projection \( P_t\) and spatial projection \( P_s\), respectively. Such information is further processed by the spatial-temporal processing to facilitate the network reuse. 

The overall INVR-M framework is shown in Fig. \ref{Fig. 2}, containing the proposed Coupled WarpRNN based multi-scale motion representation and compensation module and the mixed residual grid. Details on the proposed modules are explained in the following.

\subsection{Coupled WarpRNN based Multi-scale Motion Representation and Compensation Module}
\label{sec3.2}
\noindent To represent the regular and structured motion and appearance information in a video, a multi-scale motion representation and compensation module is proposed. A coupled WarpRNN structure is developed, based on the recurrent neural network (RNN) \cite{chung2014empirical,li2018independently,shi2015convolutional,shi2017deep,rusch2020coupled}, to efficiently represent the global and local multi-scale motion. Since RNN can naturally represent the temporal order with the recurrent connection, neither a grid or a frame index is needed as input to reconstruct a video. The information for each frame is contained in the hidden state at each step. In the following, the proposed basic WarpRNN is first presented and then the coupled WarpRNN based module is explained.

\subsubsection{\textbf {WarpRNN based Motion Representation and Compensation}}
\label{sec3.2.1}
It is known that an RNN can process the temporal information efficiently via the recurrent connection. However, the recurrent connection by adding the weight processed context focuses more on the aggregation of temporal semantic information without detailed spatial alignment. For video reconstruction, the temporal processing not only includes the overall aggregation of semantic information, but also requires the motion compensation to generate the current spatial feature based on the spatially aligned temporal feature. Therefore, a spatial warping of the temporal feature (previous context hidden state) based on the motion information is used to perform the temporal alignment first, and hence the proposed module is termed WarpRNN. Moreover, the original hidden state is also used as input to enhance the aggregation of semantic information over the whole video. In this way, the temporal context can be fully explored to produce a compensated spatial representation for each time step. In addition, to represent both the motion and appearance information, they are represented together in the neural network and obtained with motion and appearance projections when needed, similarly as the mixed grid representation as shown in (\ref{Eq.(3)}) in subsection \ref{sec3.1}.

\begin{figure}[t]
    \centering
    \includegraphics[width=0.40\textwidth]{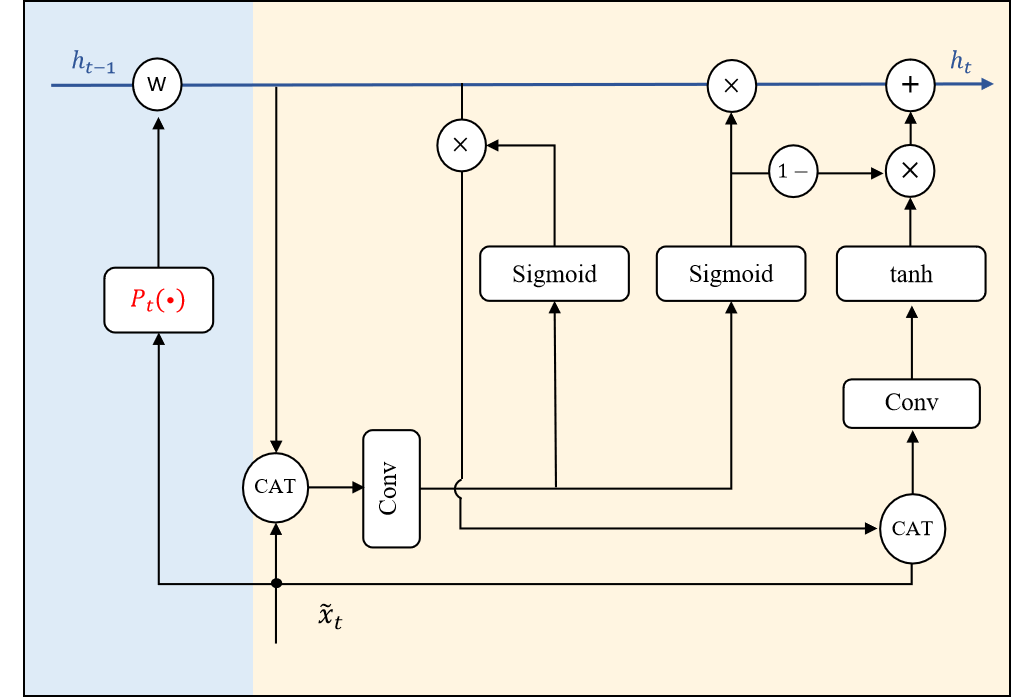}
    \caption{Illustration of the WarpRNN module, with a learned warping function.}
    \label{Fig. 3}
\end{figure}

Fig. \ref{Fig. 3} illustrates the structure of the proposed WarpRNN. For the processing at time step  \( t\) corresponding to the \( t-th\)  frame, the hidden state \( h_t\) can be obtained by
\begin{equation}
  h_t = f_g \left( \tilde{x}_t, warp\left( h_{t-1}, P_t(\tilde{x}_t) \right) \right),
  \label{Eq.(4)}
\end{equation}
where \( f_g(\cdot)\) represents the RNN network, \( h_{t-1}\) represents the hidden state at previous time step \(t-1\), and \( \tilde{x}_t\) represents the input feature, which is also \( h_{t-1}\) here to directly aggregate the temporal semantic information. 

Note that  \( \tilde{x}_t\) is further enhanced in the coupled WarpRNN shown later. \( P_t(\cdot)\) represents the projection to obtain the motion information from the hidden state, which is implemented with a convolutional layer. \( warp(\cdot)\) represents the spatial warping operation to perform the temporal alignment according to the estimated motion.

\begin{figure*}[t]
  \centering
    \includegraphics[width=0.90\textwidth]{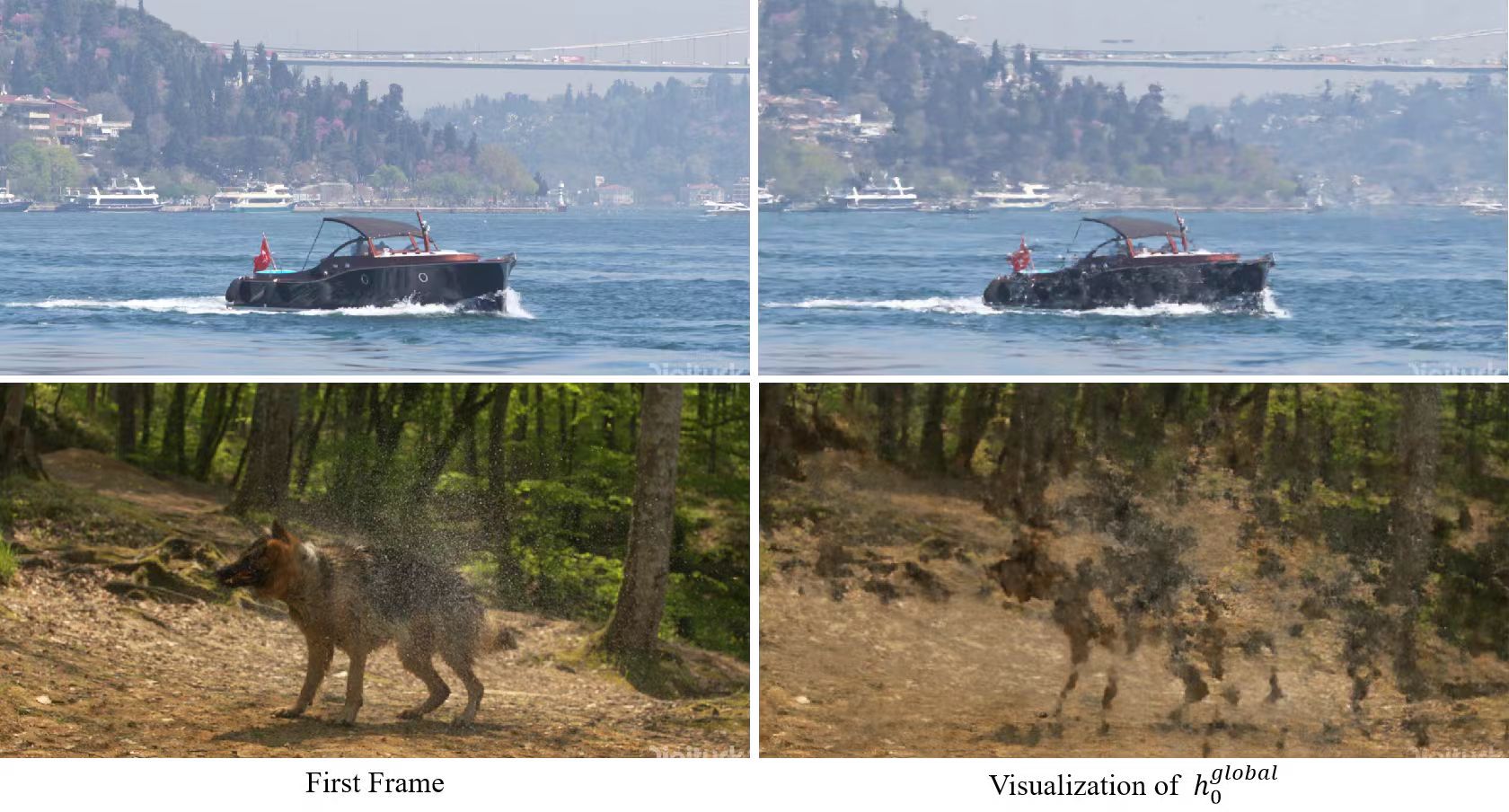}
    \caption{Illustration of the learned initial hidden state. The left column shows the ground-truth first frame. The right column shows the reconstruction result of the learnable initial hidden state $( h_0^{global})$. Note that no supervision is applied directly on the learning of $( h_0^{global})$. The information learned in $( h_0^{global})$ can be treated as an anchor frame that retains the main structure of a video and is used for predicting the following frames in the moving area. It also validates the effectiveness of the proposed method in learning the features of video frames within the hidden state of the RNN structure.} 
    \label{h0}
\end{figure*}

For the first frame without previous hidden state, a learnable tensor is used to represent an initial hidden state \( h_0\). It can be treated as an anchor frame that keeps the main structure of a video and used for predicting all the frames. The visualization result of the information contained in the initial hidden state is shown in Fig. \ref{h0}.  In the experiments, considering ConvGRU (Convolutional Gated Recurrent Unit) has fewer parameters and shown to achieve decent performance compared to the simple recurrent neural network \cite{graves2012long} and LSTM  \cite{hochreiter1997long} in processing video data, it is used as the basic RNN network \( f_g\) to construct the proposed WarpRNN.

\subsubsection{\textbf{Local and Global Motion Decomposition based Multi-scale Temporal Processing}} With the proposed WarpRNN, structured motion and appearance information in a video can be effectively represented and compensated at different time steps. However, capturing both global motions like camera shifts and local motions such as object movement within  a single network is challenging. So a local and global motion decomposition based Multi-scale temporal processing method is developed to decompose the motion information into local and global scales, which allows for adequate motion compensation of locally moving content meanwhile ensuring the motion continuity of the whole frame.

A coupled WarpRNN module is designed to serve as the multi-scale temporal processing unit. The structure of the proposed coupled WarpRNN is shown in Fig. \ref{Fig. 2}. The first WarpRNN learns to represent the local motion related information while the second WarpRNN learns the global information. The local and global motion information is decomposed by a learnable mask generated based on the feature representation at each time step. Specifically, denote the hidden states of the first and second WarpRNNs by \(h_t^{local}\) and \(h_t^{global}\), representing the learned local and global information, respectively, at time step \(t\). The hidden state \(h_{t-1}^{global}\) from the previous time step learned by the second global WarpRNN is first processed with a convolutional layer to produce a soft  mask \(W_l\) to separate local moving content from background information. Then the local content feature and background feature can be represented as
\begin{equation}
  \tilde{x}_t^{local} = h_{t-1}^{global} * W_l,
  \label{Eq.(6)}
\end{equation}
\begin{equation}
  \tilde{x}_t^{back} = h_{t-1}^{global} * (1-W_l),
  \label{Eq.(7)}
\end{equation}

For the first local WarpRNN, only the local motion related information is processed and thus the hidden state \(h_t^{local}\) at time step t can be updated as
\begin{equation}
  h_t^{local} = f_g \left( \tilde{x}_t^{local}, warp \left( h_{t-1}^{local}, P_t(\tilde{x}_t^{local})\right)\right),
  \label{Eq.(8)}
\end{equation}
where \(f_g(\cdot)\), \(warp(\cdot )\) and \(P_t(\cdot )\) follows the same definition of the proposed WarpRNN. Here it takes the local content feature generated in (\ref{Eq.(6)}) as input to focus on the local motion related information.

Then the global feature is updated by summarizing the above local information \(h_t^{local}\) and the background information \(\tilde{x}_t^{back}\) together, \(\tilde{x}_t^{global} = h_t^{local} + \tilde{x}_t^{back}\). Accordingly, the hidden state \(h_t^{global}\) of the second global WarpRNN at time step t can be updated as
\begin{equation}
  h_t^{global} = f_g \left( \tilde{x}_t^{global}, warp \left( h_{t-1}^{global}, P_t(\tilde{x}_t^{global})\right)\right),
  \label{Eq.(9)}
\end{equation}

In this way, the local and global WarpRNN can be iteratively updated in a coupled way to perform multi-scale temporal processing.  To reconstruct the  \(t-th \) frame, the global feature  \(h_t^{global}\) is first processed with a spatial projection \(P_s(\cdot) \) to generate the appearance representation feature \(F_t\) since the appearance and motion are represented together in the WarpRNN. Finally, the frame is reconstructed through a stack of NeRVBlocks, which can be represented as
\begin{equation}
\begin{aligned}
&\hat{I}_t = f_{NB}\left(P_s\left(h_t^{global}\right)\right),
\label{Eq.(10)}
\end{aligned}
\end{equation}
where  \( f_{NB}(\cdot) \) represents the NeRVBlocks. The spatial projection \(P_s(\cdot) \) can be implemented as a convolutional layer.

With this coupled WarpRNN based structure, the structured temporal motion and spatial appearance information can be efficiently represented with an explicit spatial and temporal neural network architecture. 

\subsection{Mixed Residual Grid}

\noindent While the coupled WarpRNN based neural network can provide an efficient representation for the regular and structured information in a video, relying solely on neural networks can lead to the loss of irregular information and significant performance loss as shown in Fig. \ref{Fig. 1}. It is designed to learn the missing irregular information as described in the above subsection \ref{sec3.1} Formulation. Moreover, as mentioned in Formulation, the motion and appearance are highly correlated in the lossy representation and thus modeled together as a mixed grid. Accordingly, a Mixed Residual Grid (MRG) is used to learn the irregular motion and appearance information that cannot be efficiently represented by the neural network. {Considering that it is a completely learned grid input, only the structure of the grid need be defined and the content is learned through training. }

The MRG is constructed along the temporal dimension, and for a video with \(T\) frames, it can be constructed as \(R_G \in \mathbb{R}^{L \times C \times H \times W}\), 
where \(L\), \(C\), \(H\) and \(W\) are the temporal resolution, the number of channels, height, and width of the grid, respectively. To improve the representation efficiency, \(L\) is usually set to be smaller than the sequence length \(T\). When giving the index \(t\) of a frame, the corresponding mixed residual feature \(R_t\) is temporally interpolated from \(R_G\) using bilinear interpolation. To enhance the grid representation efficiency, the proposed coupled WarpRNN is reused. Specifically, the mixed residual grid feature \( R_t  \) is first processed with a convolutional layer and then added to the hidden feature of the global WarpRNN \(h_{t-1}^{global}  \), which is then used to produce the input feature to both local and global WarpRNNs \(\tilde{x}_t^{local}  \) and \(\tilde{x}_t^{global}\), respectively, as shown in (\ref{Eq.(6)}) and (\ref{Eq.(7)}). After being compensated with the mixed residual grid feature \( R_t  \), the coupled WarpRNN can provide more detailed motion and appearance information for subsequent temporal processing and frame reconstruction.


\begin{table*}
\begin{center}
\caption{Video reconstruction results on UVG at resolution 960×1920, in terms of PSNR(dB).}
\label{tab1}
\renewcommand{\arraystretch}{1.2}
\setlength{\tabcolsep}{15pt}
\begin{tabular}{c|c c c c c c c| c   }
\hline
\hline
 \textbf{Methods} & \textbf{Beauty} & \textbf{Bosph} & \textbf{Honey}  & \textbf{Jockey} & \textbf{Ready} & \textbf{Shake} & \textbf{Yacht} & \textbf{avg}\\
\hline

    NeRV\cite{chen2021nerv}  & 33.33 &33.34&38.79&28.97&23.89&33.89&27.05&31.32 \\
    FFNeRV\cite{lee2023ffnerv}  & 33.40&33.08&38.81&31.71&24.99&33.93&27.38&31.90\\
    ENeRV\cite{li2022nerv} & 33.17&33.69&37.63&31.63&25.2&34.39&28.42&32.02\\
    DNeRV\cite{zhao2023dnerv} & 33.16&32.96&38.43&31.08&24.76&33.71&27.30&31.63 \\
    HNeRV\cite{chen2023hnerv} & 33.88&35.02&39.41&31.69&25.72&34.95&29.09&32.82 \\
    DSNeRV\cite{yan2024ds} & \underline{33.97}&\underline{35.22}&\underline{39.56}&\underline{32.86}&\underline{27.10}&\underline{35.04}&\underline{29.40}&\underline{33.31} \\
    \hline
    \textbf{Ours} &\textbf{34.69}&\textbf{35.45}&\textbf{40.47}&\textbf{33.09}&\textbf{27.15}&\textbf{35.49}&\textbf{29.78}&\textbf{33.73}\\
    
\hline 
\end{tabular}
\end{center}
\end{table*}


\begin{table*}
\begin{center}
\caption{Video reconstruction results on DAVIS at resolution 960×1920, in terms of PSNR(dB).}
\label{tab2}
\renewcommand{\arraystretch}{1.2}
\setlength{\tabcolsep}{8pt}
\begin{tabular}{c | c c c c c c c c c c | c  }
\hline
\hline
 \textbf{Methods} & \textbf{b-swan} & \textbf{b-trees} & \textbf{boat}  & \textbf{b-dance} & \textbf{camel} & \textbf{c-round} & \textbf{c-shadow} & \textbf{cows} & \textbf{dance} & \textbf{dog}  & \textbf{avg}\\
\hline

NeRV\cite{chen2021nerv}  & 25.04&25.22&30.25&25.78&23.69&24.08&25.29&22.44&25.61&27.15&25.46 \\
FFNeRV\cite{lee2023ffnerv} & 24.89&24.56&27.82&24.73&22.84&22.32&24.58&21.39&23.95&26.53&24.36\\
ENeRV\cite{li2022nerv} & 25.64&24.92&29.16&24.19&22.88&22.85&26.13&21.69&24.15&27.66&24.93\\
DNeRV\cite{zhao2023dnerv} & 29.84&28.73&30.52&26.58&26.24&28.50&28.88&24.44&28.42&30.64&28.28\\
HNeRV\cite{chen2023hnerv} & 29.23&28.67&32.27&31.39&25.93&28.72&31.21&24.67&28.43&30.72&29.12 \\
DSNeRV\cite{yan2024ds} & \underline{32.55}&\underline{29.76}&\underline{34.39}&\underline{32.21}&\underline{27.26}&\underline{29.48}&\textbf{35.88}&\underline{25.08}&\underline{28.79}&\underline{33.29}&\underline{30.87}\\
\hline
\textbf{Ours} &\textbf{33.98}&\textbf{30.06}&\textbf{34.72}&\textbf{32.79}&\textbf{27.34}&\textbf{29.62}&\underline{35.54}&\textbf{25.53}&\textbf{29.31}&\textbf{33.32}&\textbf{31.22}\\
    
\hline 
\end{tabular}
\end{center}
\end{table*}

 {The MRG can efficiently represent the irregular information compared with the neural network. The neural network and grid works together to represent the regular and structured information, and remaining irregular information, respectively. Especially for videos with large motions or scene changes, a larger grid representation can provide a better reconstruction quality. As shown in Fig. 2 with a video containing a large scene change, using the MRG greatly improves the reconstruction performance.}  The overall architecture can be learned end-to-end in the same way as the conventional INR methods. { The reconstruction loss $L_{r}$ in terms of the $L_2$ norm is used for supervision as in DSNeRV \cite{yan2024ds}. It can be represented as (\ref{lr})
\begin{equation}
  L_r = \frac{1}{T} \sum_{t=1}^{T} ||I_t - \hat{I}_t||_2,
  \label{lr}
\end{equation}
where \(   I_t  \) and \(\hat{I}_t \)  represent the ground truth and the reconstructed frame, respectively.}


\section{Experiment}
\label{sec:experiment}
\noindent Following the experimental setup of previous INVR methods \cite{chen2023hnerv,yan2024ds}, the UVG \cite{mercat2020uvg} and DAVIS \cite{perazzi2016benchmark} datasets are used for experiments.
For fair comparison, all videos from both UVG and DAVIS dataset are cropped to resolution \( 960*1920\) following HNeRV \cite{chen2023hnerv}.
Experiments including video reconstruction, video compression and video decoding are conducted for evaluation. Peak Signal-to-Noise Ratio (PSNR) is used as the metric for assessing the quality of the experiments. The Adam \cite{kingma2015adam} optimizer is used. The initial learning rate is set to $2 \times 10^{-3}$, and a cosine annealing learning rate schedule \cite{loshchilovstochastic} is used with a warm-up ratio of 0.3. 300 epochs are trained. All experiments are conducted on the Tesla V100 GPU.

\begin{figure}[t]
\setlength{\abovecaptionskip}{0.2cm}
    \centering
    \includegraphics[width=0.42\textwidth]{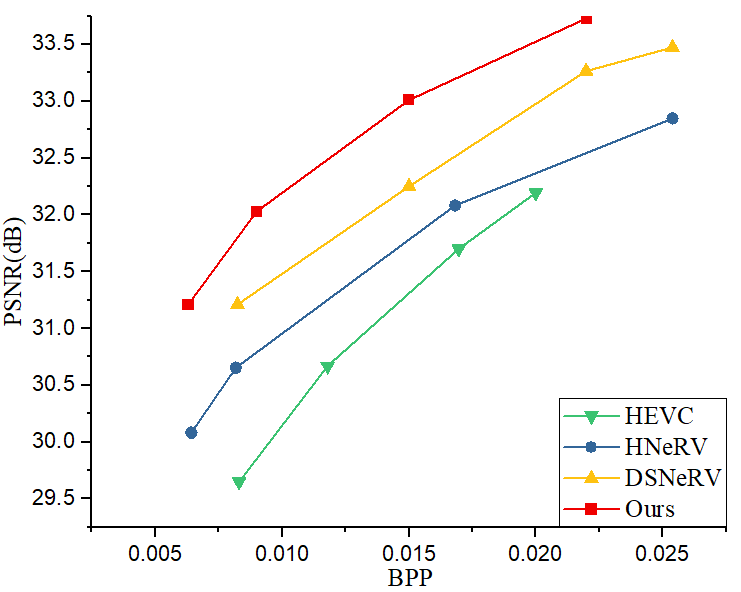}
    \caption{ R-D curve comparison on UVG dataset.}
    \label{Fig. 5}
\end{figure}

\begin{figure*}[t]
\setlength{\abovecaptionskip}{0.2cm}
  \centering
    \includegraphics[width=0.80\textwidth]{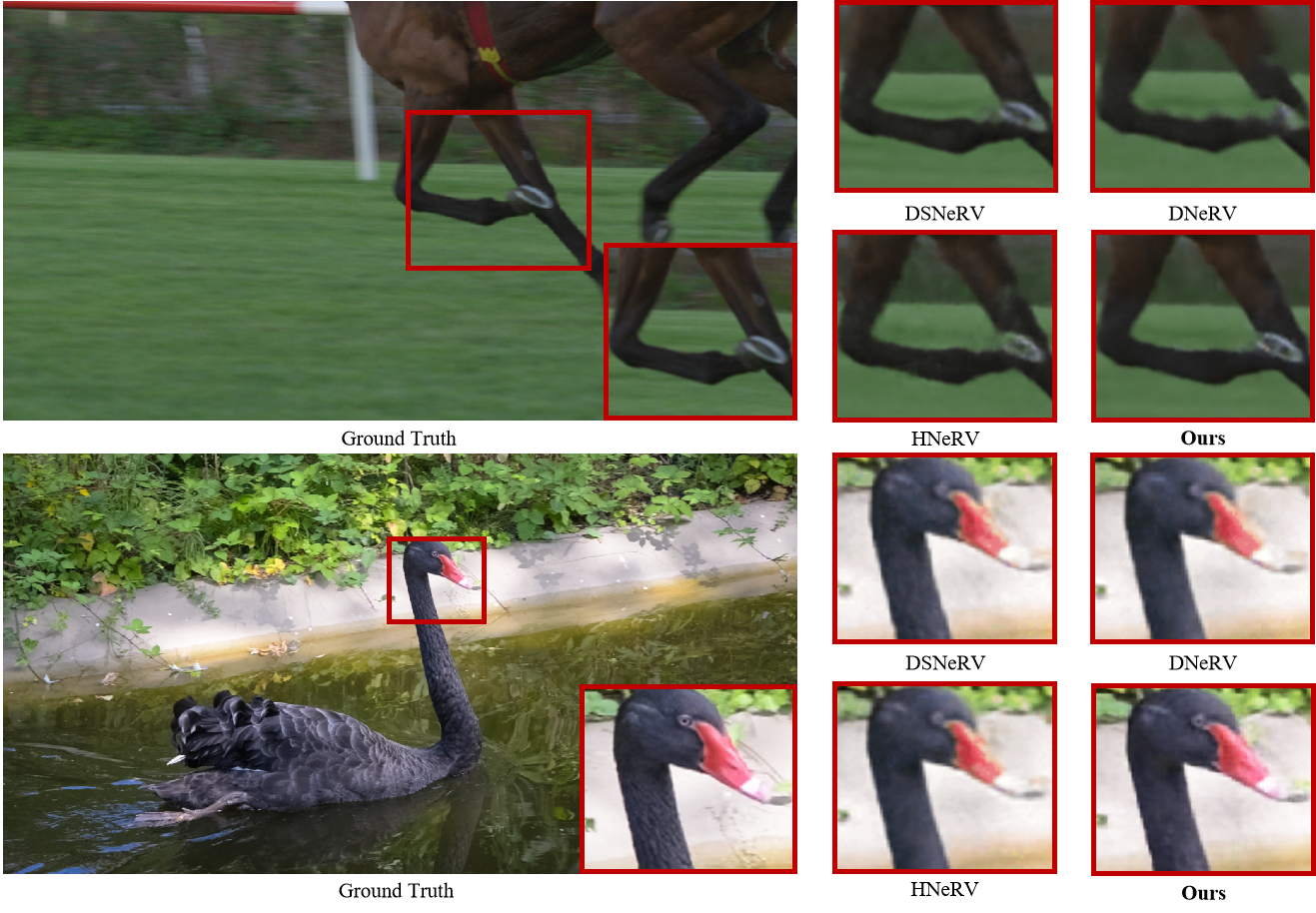}
    \caption{Example video reconstruction results on UVG and DAVIS. (Top) Jockey. (Bottom) Black swan.}
    \label{Fig. 4}
\end{figure*}

\subsection{Video Reconstruction }
\label{Video Reconstruction}
In the video reconstruction experiment, existing INVR methods including NeRV \cite{chen2021nerv}, ENeRV \cite{li2022nerv}, FFNeRV \cite{lee2023ffnerv}, HNeRV \cite{chen2023hnerv}, DNeRV \cite{zhao2023dnerv} and DSNeRV \cite{yan2024ds} are used for comparison. The model with the same model size to the compared methods, i.e., 3M parameters, is used. 

The results on the UVG dataset and DAVIS dataset are shown in Table \ref{tab1} and Table \ref{tab2}, respectively. It can be seen that our method outperforms all the existing methods in the average performance, with significant improvement over DNeRV and HNeRV.  {Moreover, it performs even better on sequences with smaller motion such as Beauty, Honey, b-swan, and etc. This actually agrees with our analysis, where the proposed method, based on the recurrent mechanism, works better at sequences with strong temporal correlation.} The qualitative results are visually illustrated in Fig. \ref{Fig. 4}. It can be seen that our method can reconstruct better quality by capturing tiny details of the frame, for example the contours of the horse hoof in motion in the jockey video and the black swan head in b-swan video. This is mainly attributed to the proposed mix residual grid module in coordination with coupled WarpRNN module, which is capable of complementing the tiny irregular residual appearance information during the reconstruction of video frames and fully exploiting the temporal motion information between video frames.

\begin{table}
\begin{center}
\caption{Compression result comparison of the proposed method against different methods on UVG dataset, in terms of BD-rate.}
\label{tab4}
\renewcommand{\arraystretch}{1.2}
\setlength{\tabcolsep}{40pt}
\begin{tabular}{c|c}
\hline
\hline
 Methods & BD-rate(\%) \\
\hline
    HEVC\cite{sullivan2012overview}     &     0 \\
    HNeRV\cite{chen2023hnerv}    &     -24.99 \\
    DSNeRV\cite{yan2024ds}   &     -31.20\\
\hline
    \textbf{Ours}     &     \textbf{-54.24}\\
\hline 
\end{tabular}
\end{center}
\end{table}

\begin{table}[t]
\begin{center}
\caption{Results of the Encoding Time and Memory Usage Comparison.}
\label{taba4}
\renewcommand{\arraystretch}{1.2}
\setlength{\tabcolsep}{15pt}
\begin{tabular}{c|c c}
\hline
\hline
 Methods & Encoding time	& Memory usage\\
\hline
    NeRV\cite{chen2021nerv}  &  3h53min  &  15.50G    \\             
    HNeRV\cite{chen2023hnerv} &  5h46min  &  11.49G    \\
    DNeRV\cite{zhao2023dnerv} &  6h11min  &  5.91G     \\
    ENeRV\cite{li2022nerv} &  3h10min  &  4.47G     \\
    FFNeRV\cite{lee2023ffnerv}&  7h43min  &  25.76G    \\
    \hline
   \textbf{Ours}  &  4h10min  &  14.22G    \\
\hline 

\end{tabular}
\end{center}
\end{table}

\subsection{Video Compression }
\label{Video Compression}
\noindent In the video compression experiment, the BD-rate is used for evaluation where the model size is transformed to bits per pixel (bpp) following the same setup as in HNeRV \cite{chen2023hnerv}. The existing video compression and INVR methods, including HEVC \cite{sullivan2012overview}, HNeRV \cite{chen2023hnerv} and DS-NeRV \cite{yan2024ds}, are used for comparison. The RD-curve result is shown in Fig. \ref{Fig. 5}. It can be seen that our method clearly outperforms the existing methods.  The BD-rate comparison is illustrated in Table \ref{tab4}, with the result of HEVC used as baseline. It can be seen that that our method can achieve a much larger bit rate saving (over 54\%) compared to other methods in the video compression task. These results confirm the efficiency of our model in video compression. 

{The encoding time and memory usage is also compared between the proposed method and the existing open-source methods. For fair comparison, all the codes are run on the same GPU platform as ours. In the training of our method, the video sequences are separated into groups, with each group fed into the RNN as a sequence. And 5 frames are used as a group in the experiments. The results are summarized in Table. \ref{taba4}. All experiments were conducted with a network size of 3M parameters on UVG dataset with 600 frames. The compared methods are trained with a batch size of 2. It can be seen that our method takes less or similar encoding time and memory usage as the existing methods except DNeRV and ENeRV. Note that our method significantly outperforms DNeRV and ENeRV with a PSNR improvement from 31.63dB and 32.02 dB to 33.73 dB on the UVG dataset.}

\begin{table}[t]
\begin{center}
\caption{Results of decoding speed, in terms of FPS.}
\label{tab5}
\renewcommand{\arraystretch}{1.2}
\setlength{\tabcolsep}{30pt}
\begin{tabular}{c|c }
\hline
\hline

 Methods &Decoding Speed\\
\hline
    {HEVC \cite{sullivan2012overview}(HM) } &    {16.56}   \\
    {HEVC \cite{sullivan2012overview}(x265)} & {152.00} \\
    NeRV\cite{chen2021nerv}  &    60.08    \\             
    HNeRV \cite{chen2023hnerv} &    62.27    \\
    DNeRV \cite{zhao2023dnerv}&   52.20     \\
    ENeRV \cite{li2022nerv}&   50.60   \\
    DSNeRV\cite{yan2024ds}&   63.54    \\
    \hline
    \textbf{Ours}  &   85.18    \\
\hline 
\end{tabular}
\end{center}
\end{table}

\begin{table}[t]
\begin{center}
\caption{Results with varying model sizes using the Big Buck Bunny dataset, in terms of PSNR(dB).}
\label{tab6}
\renewcommand{\arraystretch}{1.2}
\setlength{\tabcolsep}{18pt}
\begin{tabular}{c|c c c}
\hline
\hline
Model size & 0.75M&1.5M&3M\\
\hline
   NeRV\cite{chen2021nerv}&27.09&28.67&31.72\\
    HNeRV\cite{chen2023hnerv}&30.16&32.72&34.61\\
\hline
    \textbf{Ours}&\textbf{32.09}&\textbf{35.07}&\textbf{37.85}\\
\hline 
\end{tabular}
\end{center}
\end{table}

\begin{table}[t]
\begin{center}
\caption{Ablation study on different modules, in terms of PSNR(dB).}
\label{tab7}
\renewcommand{\arraystretch}{1.2}
\setlength{\tabcolsep}{5pt}
\begin{tabular}{c|c |c| c | c  c c  }
\hline
\hline
\  &      &Single&Coupled&\ &&\\
\  &  MRG &WarpRNN& WarpRNN&\ Bunny&Beauty&Honey\\
\hline
    v1&  & & \checkmark&37.28&33.61& 38.66\\
\hline
   v2&\checkmark  & \checkmark  & &37.69&33.50&39.32\\

\hline
    v3& \checkmark & & \checkmark  &37.85&34.69&40.47\\
\hline 
\end{tabular}
\end{center}
\end{table}

\begin{figure}[t]
    \centering
    \includegraphics[width=0.48\textwidth]{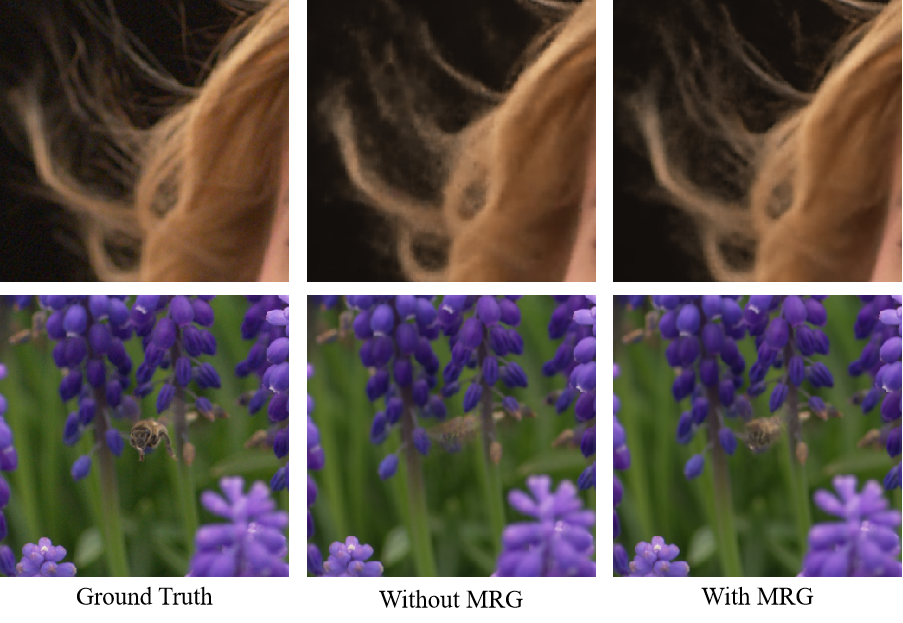}
    \caption{Example qualitative result comparison on the mixed residual grid.}
    \label{Fig. 6}
\end{figure}

\subsection{Video Decoding}
\label{Video Decoding}
\noindent The video decoding speed is an important factor in practical use, and the decoding speed of our method is also evaluated and compared with HEVC \cite{sullivan2012overview}, NeRV \cite{chen2021nerv}, HNeRV \cite{chen2023hnerv}, DNeRV \cite{zhao2023dnerv}, ENeRV \cite{li2022nerv} and DSNeRV \cite{yan2024ds}. The results are presented in Table \ref{tab5}, where the decoding speed is measured in FPS (frames per second). {Two HEVC implementations are used, including  the HEVC Test Model (HM) and the x265, and both are run on CPU. }All INVR methods are run on a single GPU on Tesla V100. It can be seen that our method not only achieves outstanding reconstruction performance but also decodes faster than other INVR methods, which is of great significance in some application scenarios.

\subsection{Ablation Studies }
\label{Ablation Studies}
\noindent The Big Buck Bunny dataset \cite{roosendaal2008big} is used in the ablation experiments. And two videos from UVG dataset (“Beauty” and “Honey”) are further used for evaluating the proposed Coupled WarpRNN and Mixed Residual Grid.

\noindent\textbf{Varying Model Size.} An ablation experiment on different model sizes is first conducted. The NeRV \cite{chen2021nerv} and HNeRV \cite{chen2023hnerv} are also added for comparison as representative INVR-N method and INVR-G method, respectively. The results are shown in Table \ref{tab6}. It can be seen that our method outperforms both methods in terms of reconstruction quality at all sizes, demonstrating the representational capability of our method across different model sizes.

\noindent\textbf{Coupled WarpRNN.} The experiments of only using the proposed coupled WarpRNN without grid input, single-layer ConvGRU, single-layer WarpRNN and coupled WarpRNN with grid (with same model size 3M) are all conducted to validate the effectiveness of the proposed Coupled WarpRNN. The results are shown in Table \ref{tab7}. It can be seen that only using the proposed coupled WarpRNN without grid (v1 in Table \ref{tab7}) can also work with a decent performance, indicating that it can effectively represent most of the regular and structure information in a video. Then by comparing the performance of using single WarpRNN and coupled WarpRNN (v2 and v3 in Table \ref{tab7}), it can be seen that the proposed coupled WarpRNN can significantly improve the performance around 1dB for the camera captured videos. For the Big Buck Bunny video, it can still improve the performance although not significant.


\noindent\textbf{Mixed Residual Grid (MRG). }The effectiveness of the mixed residual grid can be observed by comparing the performance of v1 and v3 in Table \ref{tab7}, representing the results without and with the mixed residual grid, respectively. It can be seen that the model with the mixed residual grid performs better than the model without it.  Some example qualitative result comparisons are shown in Fig. \ref{Fig. 6}. It can be seen that the results with mixed residual contain more details such as the hair and bee, indicating that the mixed residual grid can learn the irregular  information that cannot be learned solely by neural networks to improve video reconstruction quality.

A visual comparison of using the Mixed Residual Grid is also conducted as shown in Fig. \ref{motivation} in both spatial and temporal domains. In addition to better represent the spatial details, the grid can also better represent the temporal irregular information such as sudden scene change. To better illustrate this, a new video created by inserting n frames of the Beauty video between 2N frames of a blooming lily video is used for reconstruction. When there are changes in the temporal domain of video frames, MRG can effectively capture these sudden variations as irregular information, thereby improving the reconstruction quality of video frames. This further validates the importance of combining the neural network and MRG.

\section{Conclusion}
\label{Conclusion}
\noindent In this paper, we propose a coupled WarpRNN based implicit neural video representation method, CWRNN-INVR, based on the INVR-M framework. The roles of neural network and grid in INVR are first thoroughly investigated, and demonstrates that neural network can better represent structured information with unified representation while grid can better represent irregular information with locally learned representation. This motivates the proposed INVR-M framework by decomposing a video into regular and structured information and irregular information. The Coupled WarpRNN based multi-scale motion representation and compensation module further incorporates the temporal processing into neural network based on global and local motion decomposition. The mixed residual grid represents the remaining motion and appearance information together with network reuse. Experimental results show that the proposed method achieves better performance than the existing INVR methods and ablation study verifies the effectiveness of each proposed module.


%

\end{document}